\documentclass[final] {aipproc}
\usepackage{sidecap,graphicx,boxedminipage,epsfig,wrapfig,floatflt,shadow}

\layoutstyle{8x11double}

\begin{document}

\title{Coupling Conceptual and Quantitative Problems to Develop Expertise in Introductory Physics Students
\classification{01.40Fk,01.40.gb,01.40G-,1.30.Rr}
\keywords      {problem solving, introductory physics, conceptual learning, quantitative problems}
}

\author{Chandralekha Singh}{
  address={Department of Physics and Astronomy, University of Pittsburgh, Pittsburgh, PA, 15260}}

\begin{abstract}
We discuss the effect of administering conceptual and quantitative isomorphic
problem pairs (CQIPP) back to back vs. asking students to solve only one of the problems in the CQIPP
in introductory physics courses. Students who answered both questions
in a CQIPP often performed better on the conceptual questions than
those who answered the corresponding conceptual questions only. Although
students often took advantage of the quantitative counterpart to answer
a conceptual question of a CQIPP correctly, when only given the conceptual
question, students seldom tried to convert it into a quantitative
question, solve it and then reason about the solution conceptually.
Even in individual interviews, when students who were only given conceptual
questions had difficulty and the interviewer explicitly encouraged
them to convert the conceptual question into the corresponding quantitative
problem by choosing appropriate variables, a majority of students
were reluctant and preferred to guess the answer to the conceptual
question based upon their gut feeling. 
\end{abstract}
 
\maketitle

\section{Introduction}

We investigate whether introductory physics students perform better when they answer a conceptual and quantitative isomorphic
problem pair (CQIPP) as a set compared to the case when they answer either the conceptual or the quantitative question of a CQIPP alone. 
We call the paired problems isomorphic if they require the same physics principle to solve them. 
Although it is difficult to categorize physics questions as exclusively quantitative or conceptual, CQIPPs had one
question that required symbolic or numerical calculation while the
other question could be answered by conceptual reasoning alone~\cite{singh}. 
We also analyze the performance of students on the CQIPPs from the perspective of {}``transfer\char`\"{}~\cite{mestre,sanjay}.

We developed five CQIPPs in the multiple-choice format with different contexts in mechanics~\cite{singh}.   
Below is an example of the first CQIPP (correct answers in italics):

(1) A tugboat pulls a ship of mass $M$ into the harbor with a constant tension force $\vec F$ in
the horizontal tow cable. Both the tugboat and the ship start from rest. After the ship has been towed
a distance $d$ in time $t$, the magnitude of its momentum will be\\
(a) $Fd$\\
(b) $(1/2) (F/M)t^2$\\
(c) $(F/M) t^2/d$\\
(d) $(1/2) (F/M) d t^2$\\
{{\it (e) $F t$}}\\

\noindent
(2) Two identical tugboats pull other ships as shown below, starting from rest. The Queen
Mary is a much more massive ship than the Minnow. Both tugboats pull with the same horizontal force.
Neglect other forces. After both tugboats have been pulling for the same amount of time, which
one of the following is true about the Queen Mary and the Minnow?\\
\begin{center}
\epsfig{file=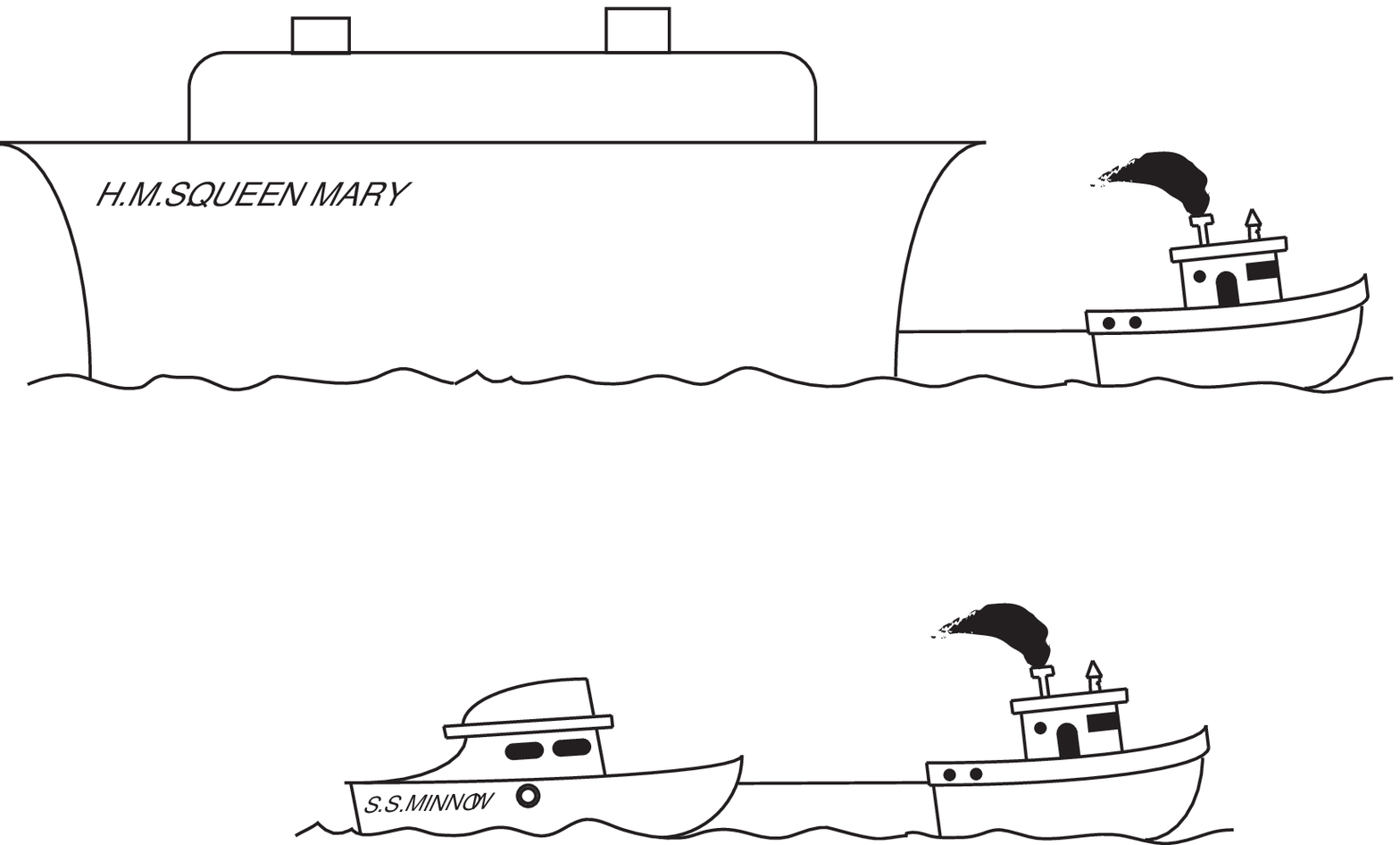,height=.9in}
\end{center}
(a) The Queen Mary will have a greater magnitude of momentum.\\
(b) The Minnow wil have a greater magnitude of momentum.\\
{{\it (c) Both ships will have the same magnitude of momentum.}}\\
(d) Both ships will have the same kinetic energy.\\
(e) The Queen Mary will have a greater kinetic energy.\\

We made hypotheses H1 and H2 as described below:
%\begin{itemize}
%CQIPPs in which one question is more quantitative than the other: 
\begin{itemize}
\item H1: 
Performance on quantitative questions of a CQIPP will be better when both the quantitative and conceptual questions
are given than when only the quantitative question is given.
\item H2: 
Performance on conceptual questions of a CQIPP will be better when both the quantitative and conceptual questions
are given than when only the conceptual question is given.

\end{itemize}

Hypothesis H1 is based on the assumption that solving the conceptual question of a CQIPP may 
encourage students to perform a qualitative analysis, streamline students' thinking, 
make it easier for them to narrow down relevant concepts, and thus help them solve the quantitative problem correctly.
Prior studies show that introductory physics students are not systematic in using effective problem solving strategies, and often do not perform
a conceptual analysis while solving a quantitative problem. 
They often use a ``plug and chug" approach to solving quantitative problems, which may prevent them from solving the problem correctly.
The conceptual questions may provide an opportunity for reflecting upon the quantitative problem and performing a qualitative analysis and planning.
This can increase the probability of solving the quantitative problem correctly.
We note that since the CQIPPs always had a quantitative question preceding the corresponding conceptual question, hypothesis
H1 assumes that students will go back to the quantitative question if they got some insight from the corresponding conceptual question.

Hypothesis H2 is inspired by prior studies that show that introductory physics students often perform 
better on quantitative problems compared to conceptual questions on the same topic~\cite{lillian}.
Students often treat conceptual questions as guessing tasks~\cite{lillian}.
We hypothesized that students who are able to solve the quantitative problem in a CQIPP may use its solution as a hint for answering
the conceptual question correctly if they are able to discern the similarity between the two questions.
Since quantitative and conceptual questions of a CQIPP were given one after another,
we hypothesized that students would likely discern their underlying similarity at least in cases where the contexts were similar. 
When reasoning without quantitative tools, it may be more difficult to create the correct chain of reasoning
if a student is ``rusty" about a concept~\cite{fred2}. Equations can provide a pivot point for constructing the reasoning chain.
For example, if a student has forgotten whether or not the maximum safe driving speed while making a turn
on a curved road depends on the mass of the vehicle, he/she will have great difficulty
determining which is correct without using equations to reason.
Similarly, a student with evolving expertise who is comfortable reasoning with equations may need to
write down Newton's second law {\it explicitly}
to conclude that the tension in the cable of an elevator accelerating upward is greater than its weight.
An expert can use the same law {\it implicitly} and conceptually argue that the upward acceleration implies that
the tension exceeds the weight without writing down Newton's second law explicitly.
Being able to reason conceptually without resorting to quantitative tools in a wide variety of contexts
may be a sign of adaptive expertise whereas conceptual reasoning by resorting to quantitative tools
may be a sign of evolving expertise~\cite{fred2}.

Students in college calculus-based introductory physics courses participated in the study.
The questions were asked after instruction in relevant concepts and after students
had an opportunity to work on their homework on related topics. 
When students were given both questions of a CQIPP back to back, 
the quantitative questions preceded the corresponding conceptual question. 
However, students were free to go back and forth between them if they wished and could change
the answer to the previous question if they acquired additional insight
for solving the previous question by answering a latter question.
Students who were given both questions of a CQIPP were \underline{not} told
explicitly that the questions given were isomorphic. They were given
2.5 minutes on an average to answer each question.
In some of the courses, we discussed the responses individually with several student volunteers. 

\vspace*{-0.1in}
 
\section{Results and Discussion}

\vspace*{-0.1in}

Table 1 summarizes the numbers of students who were given
both or one question of a CQIPP, and students' average performance. 
Table 1 also shows the results of a Chi-square test with $p$ values for comparison between
cases when both questions of a CQIPP were given vs. only one question was given.
Students can make appropriate connections between the questions in a CQIPP only if they
have a certain level of expertise that helps them discern the connection
between the isomorphic questions. Improved student performance when both
questions of a CQIPP were given vs. when only one question
was given was taken as one measure of transfer of relevant knowledge from one problem to another.
\begin{table}[h]
\centering
\begin{tabular}[t]{|c|c|c|c|}
\hline
Problem $\#$ & only one & both & p value\\[0.5 ex]
\hline
1& 59 (138)&54 (289)& 0.40 \\[0.5 ex]
2& 31 (215) & 58 (289)&0.00\\[0.5 ex]
\hline
3& 34 (138) &38 (289)& 0.45 \\[0.5 ex]
4& 23 (215) &30 (289) &0.07 \\[0.5 ex]
\hline
5& 81 (138) &76 (289) &0.26\\[0.5 ex]
6& 55 (215) &80 (289) &0.00\\[0.5 ex]
\hline
7& 52 (138) &56 (289)& 0.47 \\[0.5 ex]
8& 44 (150) & 51 (289)& 0.19 \\[0.5 ex]
\hline
9 & 49 (138)& 49 (289) & 1.00 \\[0.5 ex]
10 & 53 (150)& 71 (289) & 0.00 \\[0.5 ex]
\hline
\end{tabular}
\vspace{0.1in}
\caption{For the CQIPP given, the first column lists
the problem numbers, the second column gives the percentage of students who chose the correct answer when
only one of the questions was given to them, and the third column gives the percentage of students who chose
the correct answer when both questions were given. 
The numbers in parentheses in the second and third columns refer to the number
of students who answered the question. The last column for all questions 
lists the p value for comparison of student performance between cases when only one of the 
questions in a CQIPP was given vs. when both questions were given.
}
\label{junk}

\end{table}

Table 1 shows that, contrary to our hypothesis H1, student performance on quantitative questions was not significantly different 
when both quantitative and conceptual questions were given back to back (with the quantitative question
preceding the conceptual question) than when only the corresponding quantitative question was given. 
In some cases, the performance on the conceptual question was better than the performance on the quantitative
question (problem pairs (9)-(10)), but students could not leverage their conceptual knowledge for gain on the corresponding
quantitative problem. As noted earlier, the two questions in a CQIPP were
always given in the same order, although students could go back and forth
if they wanted. It is possible that students did not go back to
the questions they had already answered, especially due to the time constraint, even if the question that followed
provided a hint for it. Future research will evaluate the effect of
switching the order of the quantitative and conceptual questions in a CQIPP when both are given. 

On the other hand, in support of hypothesis H2, students who worked on both questions of the IPPs
involving a conceptual and a quantitative problem performed better
on the conceptual questions at least for three of the five IPPs than when they were given only the conceptual questions. 
Table 1 shows that, for three of the IPPs, students often performed better on the conceptual question
when both questions were given rather than the corresponding conceptual question alone.
The fact that many students took advantage of the quantitative problem to solve the conceptual question points to their evolving expertise. 
For example, many students who were given both questions (1) and (2) recognized that the final momenta of the ships 
are independent of their masses under the given conditions by solving the quantitative problem. 
Written responses and individual discussions suggest that some students who
answered the conceptual question (2) correctly were not completely
sure about whether the change in momentum in question (1) was given
by option (a) or (e). However, since the answer in either case is
independent of the mass of the object, these students chose the correct
option (c) for question (2). The students who chose the incorrect
option (a) for question (1) but the correct option (c) for question
(2) often assumed that both ships in question (2) must have traveled
the same distance although that is not correct. 

In the third CQIPP, question (5) asks students to calculate (numerically) the speed of a person at
the bottom of a $5$ m high slide if the person started from rest
on the top. Question (6) asks students to compare the speeds of two
people with different masses at the bottom of the slide, who started
from rest at the top of the slide.
In individual discussions, several students explicitly noted that
the mass cancels out in question (5) so the answer to question (6) cannot depend on mass. 
In the last CQIPP pairing questions (9) and (10), question (9) asks students
to numerically calculate the final speed of a boat moving horizontally
when Batman falls vertically into it and comes to rest with respect
to the boat. Question (10) asks a conceptual question about the final
speed of a cart moving horizontally when rain falls vertically into
the cart and comes to rest and also asks students about the physics
principle involved in arriving at the solution. 
Discussions with individual students and students' written work suggest that solving the
quantitative question (9) helped many students formulate their solution to question (10). 
Although some students were not able to solve the
quantitative question, e.g., due to algebraic error or not realizing
that when considering the conservation of the horizontal component of
momentum, Batman's vertical velocity should not be included, it was
easier for them to answer the conceptual question after thinking about
the quantitative one. Most realized that the boat would slow down after Batman lands in it. 

Previous research shows that answering conceptual questions can sometimes be more challenging for students than quantitative
ones, if the quantitative problems can be solved algorithmically and
students' preparation is sufficient to perform the mathematical manipulations~\cite{lillian}.
If a student knows which equations are involved in solving a quantitative
problem or how to find the equations,
he or she can combine them in \textit{any order} to solve
for the desired variables even without a deep conceptual understanding
of relevant concepts. On the contrary, while reasoning without equations,
the student must usually proceed in a \textit{particular order} in
the reasoning chain to arrive at the correct conclusion~\cite{lillian}.
Therefore, the probability of deviating from the correct reasoning
chain increases rapidly as the chain becomes long. We note however
that our hypothesis H2 is not about whether students will perform better
on the quantitative or conceptual question of a CQIPP when the
two questions are given separately (especially because the wording is not parallel for the
quantitative and conceptual questions in a CQIPP). Rather, our hypothesis
relates to whether students will recognize the similarity of the quantitative
and conceptual questions in a CQIPP, and take advantage of their solution to one question to answer the corresponding paired question.
Our finding suggests that students can leverage their quantitative solutions to correctly answer the corresponding
conceptual questions, at least in the questions given.

The fact that students often performed better on conceptual questions
when they were paired with quantitative questions brings up the following
issue. If students could turn the conceptual questions into analogous
quantitative problems themselves when only the conceptual questions
were given, they may have solved the quantitative problem algorithmically
if they were comfortable with the level of mathematics needed, and
then reasoned qualitatively about their results to answer the original
conceptual question. Almost without exception, students did not do this in the interview situation.
One can hypothesize that students have not thought seriously about
the fact that a conceptual question can be turned into a quantitative
problem, or that a mathematical solution can provide a tool for reasoning
conceptually. Without explicit guidance, students may not realize
that this conversion route may be more productive than carrying out
long conceptual reasoning without mathematical relations. However,
we find that students avoided turning conceptual questions into quantitative
ones, even when explicitly encouraged to do so. In one-on-one interview
situations, when students were only given the conceptual questions,
they also tried to guess the answer based upon their gut feeling.
More research is required to understand why students are reluctant
to transform a conceptual question into a quantitative problem even
if the mathematical manipulations required after such a conversion
and making correct conceptual inferences are not too difficult for them. One
possible explanation for such reluctance is that such a transformation
from a conceptual to a quantitative problem is cognitively demanding for
a typical introductory physics student and may cause a mental overload~\cite{sweller}.
According to Simon's theory of bounded rationality, an individual's
rationality in a particular context is constrained by his/her expertise
and experience, and an individual will only choose one of the few options
consistent with his/her expertise that does not cause a cognitive overload.~\cite{simon}

In the second CQIPP, problem (3) asks students to calculate the speed of a hoop rolling down a ramp
given the various parameters for the hoop and the ramp. Problem (4)
asks them to compare the speeds of two different hoops with different masses and radii rolling down the same ramp. 
For this CQIPP, the quantitative problem was very challenging.
Most interviewed students and those who wrote something on their answer sheet did not
use conservation of energy correctly and forgot to take into account
both the rotational and translational kinetic energies in their analysis.
Thus, it is not surprising that there is no significant difference between cases when only one
of the questions was given vs. when both were given.

\vspace*{-0.1in}
 
\section{Conclusions}

While students often took advantage of the quantitative problem to answer the corresponding
conceptual question of a CQIPP, those who were only given the corresponding conceptual
question did not automatically convert it into a quantitative problem as an aid for reasoning correctly.
Examination of students' scratch work suggests that they seldom attempted such conversion 
by choosing appropriate variables.
One-on-one discussions suggest that students often used gut feeling to reason about the conceptual questions.
This tendency persisted even when the interviewer explicitly encouraged students to convert a conceptual question into
a quantitative one. It is possible that converting the conceptual questions to quantitative ones was too cognitively 
demanding for students and may have caused mental overload.

In this research, isomorphic problems were given back-to-back, and
the more quantitative question always preceded the conceptual question in a CQIPP. 
It is possible that the order in which questions were asked and
the proximity of the paired questions in a CQIPP are major factors in whether students 
will recognize their similarity and transfer relevant knowledge from one problem to another. 
In future research, one can explore the effect of
spacing the questions in a CQIPP and changing the order in which questions are asked
on students' ability to benefit from having both questions of a CQIPP.

Presenting quantitative and conceptual isomorphic pairs
helped students make conceptual inferences using quantitative tools.
Such problem pairs as part of instruction may help students go beyond
the {}``plug and chug\char`\"{} strategy for the quantitative
problem solving and may give them an opportunity to reflect upon their
solution and develop reasoning and meta-cognitive skills. Solving
these paired problems can force students to reflect upon the problem
solving process and improve their meta-cognitive skills. Helping students
develop meta-cognitive skills can also improve transfer of relevant knowledge from one problem to another.

\begin{theacknowledgments}

We thank 
NSF for award DUE-0442087.
\end{theacknowledgments}
%\vspace*{-.10in}

\bibliographystyle{aipproc}

%\pagebreak

%\input{ntable2}

\end{document}